\documentclass[prl,twocolumn,showpacs,amsmath,amssymb,superscriptaddress]{revtex4}
\usepackage{graphicx}
\usepackage{dcolumn}
\usepackage[sort&compress]{natbib}
\usepackage{subfigure}
\usepackage{ifpdf}
\usepackage{bm}
\ifpdf
\usepackage[pdftex,
        colorlinks=true,
        pdftitle={},
        pdfauthor={S. Salem-Sugui Jr.},
        pdfsubject={phase coherence},
        pdfkeywords={IF, UFRJ, Rio de Janeiro Brazil},
        baseurl={http://www.if.ufrj.br},
        pdfpagemode=UseNone,
        bookmarksopen=true
        pdfoutput=1
        ]{hyperref}
\fi

\begin{document}
\title{Onset of phase correlations in YBa$_2$Cu$_3$O$_{7-x}$ as determined from reversible magnetization measurements}
\author{S. Salem-Sugui Jr.}
\affiliation{Instituto de Fisica, Universidade Federal do Rio de Janeiro,
21941-972 Rio de Janeiro, RJ, Brazil}
\author{A. D. Alvarenga}
\affiliation{Instituto Nacional de Metrologia Normaliza\c{c}\~ao e
Qualidade Industrial, 25250-020 Duque de Caxias, RJ, Brazil.}
\date{\today}
\begin{abstract}

Isofield magnetization curves are obtained and analyzed for three single crystals of YBa$_2$Cu$_3$O$_{7-x}$, ranging from optimally doped to very underdoped, as well as the BCS superconductor Nb, in the presence of magnetic fields applied both parallel and perpendicular to the $ab$ planes.  Near $T_c$, the magnetization exhibits a temperature dependence $\sqrt{M}\propto [T_a(H)-T]^m$.  In accordance with recent theories, we associated $T_a(H)$ with the onset of coherent phase fluctuations of the superconducting order parameter.  For Nb and optimally doped YBaCuO, $T_a(H)$ is essentially identical to the mean-field transition line $T_c(H)$.  The fitting exponent $m \simeq 0.5$ takes its mean-field value for Nb, and varies just slightly from 0.5 for optimally doped YBaCuO.  However, underdoped YBCO samples exhibit anomalous behavior, with $T_a(H)>T_c$ for $H\parallel c$, suggesting that the magnetization is probing a region of temperatures above $T_c$ where phase correlations persist.  In this region, the fitting exponent falls in the range $0.5 < m < 0.8$ for $H\parallel c$, compared with $m\simeq0.5$ for $H\parallel ab$.  The results are interpreted in terms of an anisotropic pairing symmetry of the order parameter: $d$-wave along the $ab$ planes and $s$-wave along the $\hat{c}$ axis. 
\end{abstract}
\maketitle 

\section{Introduction}
It is well known that underdoped high-$T_c$ superconductors exhibit a rich variety of fluctuation phenomena in the pseudo-gap regime \cite{timusk,lee}.  Many of these phenomena are thought to arise from amplitude fluctuations, which persist even without  phase coherence \cite{wang1,wang2}.  However, the superconducting transition at zero magnetic field, $T_c$, represents the temperature at which phase coherence is lost \cite{emery}.  For low-$T_c$ superconductors, neither amplitude nor phase fluctuations play an important role, and the resulting transition is of the mean-field type \cite{emery}.  On the other hand, high-$T_c$ materials, as underdoped cuprates, have a low superfluid density \cite{emery}, leading to a small phase stiffness and an enhancement of phase fluctuations in the vicinity of $T_c$.   In Ref.~[\onlinecite{beck}], a line of phase transitions $T_{\phi}(x)>T_c$ was obtained for YBa$_2$Cu$_3$O$_{7-x}$ as a function of doping $x$, marking the disappearance of  phase coherence.  However, Refs.~[\onlinecite{emery}] and [\onlinecite{beck}] did not consider an applied magnetic field, although most superconducting phenomena above $T_c$ in underdoped materials depend sensitively on the magnetic field \cite{lee,wang1,wang2,said} and fluctuating vortices.  The dearth of experimental work in the literature, exploring the existence of the temperature $T_{\phi}$ , forms the motivation for the present experiments.  As we shall explain below, our work was also motivated by the results of Ref.~[\onlinecite{kwon}], where phase fluctuations are shown to have an effect on the superfluid density of states, reducing the gap in the vicinity of $T_c$. 

In this paper, we present isofield measurements of the reversible magnetization versus temperature, obtained from three single crystals of YBaCuO ranging from underdoped to optimally doped.  We consider magnetic fields applied both parallel and perpendicular to the $\hat{c}$-axis.  We also present a few isofield $\sqrt{M}$ vs. $T$ curves for Nb, a well know BCS superconductor \cite{giaever}, for comparison.  Our data are plotted as $\sqrt{M}$ vs.\ $T$, where $\sqrt{M}$ is directly proportional to the amplitude of the order parameter $|\psi|$ near $T_c$ \cite{deGennes}.  In this way, we can directly probe fluctuations of the order parameter occurring below and above $T_c$ in YBaCuO within the pseudo-gap region \cite{ding,timusk,lee}. 

\section{Experimental}
The single crystals of YBaCuO were grown at Argonne National Laboratory \cite{veal} with fully developed
transitions of width $\Delta T_{c}\simeq 1-2K$.  The optimally doped $YBa_2Cu_3O_{7-x}$ single crystal had a composition of $x\sim 0.05$ with $T_c$=91.7 K , and have approximate dimensions of 1x1x0.3 mm and mass $\simeq 1.7 mg$ .  For the underdoped samples with $H\parallel \hat{c}$, we used magnetization data previously obtained from two deoxygenated crystals \cite{JLTP1} with compositions of $x=0.5$ and 0.6, and corresponding transitions at $T_c$=41.5 and 52~K, respectively (these two samples have approximate dimensions of 1x1x0.2 mm and mass $\simeq 1 mg$) .  For the Nb sample, we used the magnetization data previously presented in Ref.~[\onlinecite{saidnb}].  All our magnetization data were obtained with a commercial magnetometer, based on a superconducting quantum interference device (SQUID).  The data were obtained after cooling the sample from temperatures above $T_c$ in zero applied magnetic field to a desired temperature below $T_c$.  Magnetic fields up to 50~kOe were then applied, reaching the desired value without overshoot.  The data were obtained by heating the sample in fixed increments of temperature up to a temperature well above $T_c$.  We also obtained field-cooled curves, which allowed to obtain the reversible (equilibrium) magnetization.  Two field orientations were used, corresponding to $H\parallel \hat{c}$ and $H\parallel ab$ planes.  For the latter geometry, we attached the samples with varnish to a tiny quartz-rod inserted along the axis of a straw.  The magnetization data were obtained up to 150~K for the $T_c$=91.7~K sample, up to 80~K for the $T_c$=52~K sample, and up to 70~K for the $T_c$=41.5~K.  The background magnetization due to normal state electrons was determined (and removed) for each data set by fitting to the form $M_{b}=c(H)/T-a(H)$ in a temperature window well above $T_c$. 

\section{Results}
\begin{figure}[t]
\includegraphics[width=\linewidth]{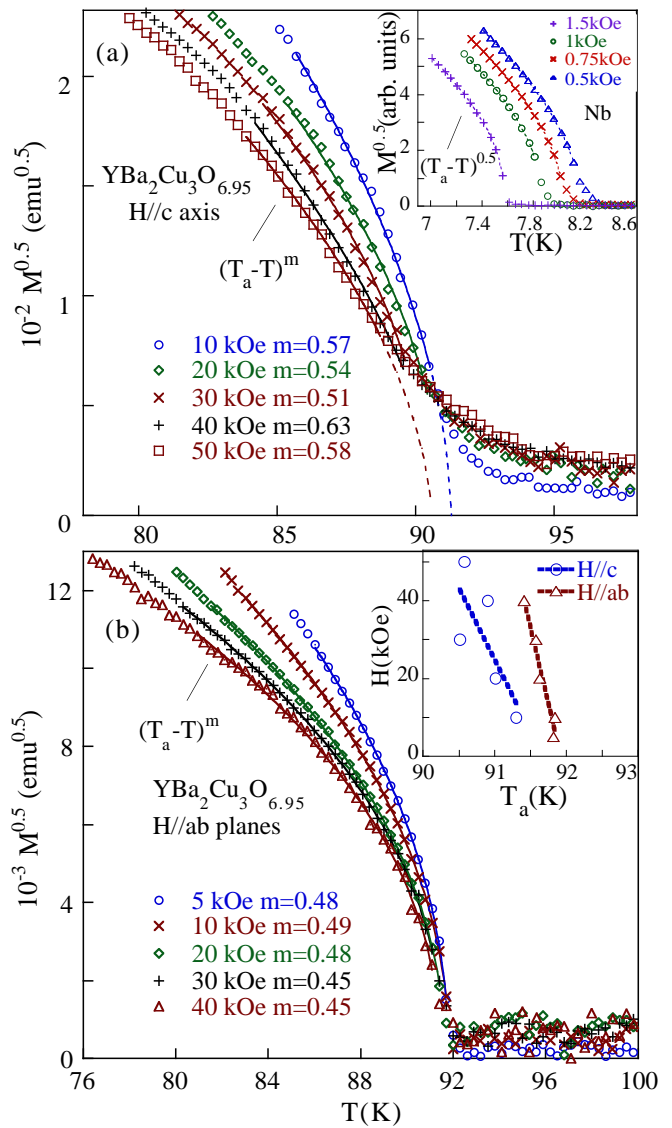}
\caption{Isofield curves of $\sqrt{M}$ vs.\ $T$ where $M$ is the reversible magnetization. (a) Optimally-doped YBaCuO ($T_c$=91.7~K) with $H\parallel c$. Dashed  lines shown the extrapolation of the fittings to $T_a(H)$ for $H=$ 50 and 10 kOe. Inset:  Nb data (note that Nb is isotropic).  (b) Optimally-doped YBaCuO with $H\parallel ab$.  Inset:  phase correlation temperature $T_a(H)$ vs.\ $H$.}
\label{92K}
\end{figure}

The resulting reversible magnetization data are plotted as $\sqrt{M}$ vs.\ $T$ in Figs.~\ref{92K}, \ref{52K} and \ref{41K}.  The Nb data are shown in the inset of Fig.~\ref{92K}(a).  As a reference point, the $H\parallel \hat{c}$ data, shown in Figs.~\ref{92K}(a), \ref{52K}(a), and \ref{41K}(a), all exhibit a crossing point, where the magnetization is independent of the field.  This interesting feature occurs in 2D systems \cite{bula,tesa,rosenstein} as well in 3D systems \cite{zac,rosenstein}, and can be explained in terms of vortex fluctuations.  The crossing point was previously investigated for the same deoxygenated samples studied here \cite{JLTP1}, and it will not be discussed further.  

In the conventional theory of the upper critical field $H_{c2}$ \cite{abrikosov}, the magnetic induction $B$ obtained from the Ginzburg-Landau equation can be expressed as \cite{deGennes}
\begin{equation} 
B=H-\frac{4\pi e \hbar}{mc}|\psi|^2 ,
\end{equation}
where $\psi$ is the superconducting order parameter. The magnetisation $M=(B-H)/4\pi$ is then given by
\begin{equation} 
M=-\frac{e \hbar}{mc}|\psi|^2 .
\end{equation}
Within the Abrikosov approximation \cite{abrikosov}, it therefore follows that $\sqrt{M}$ is directly proportional to the average amplitude of the order parameter.  Near the superconducting transition, the temperature dependence of the magnetization can then be expressed as $\sqrt{M} \propto [T_c(H)-T]^m$, in terms of the mean-field transition temperature $T_c(H)$.  The mean-field exponent is given by $m=1/2$ both for $s$-wave BCS superconductors \cite{deGennes}, and for $d$-wave superconductors within a Ginzburg-Landau theory \cite{xu}. 

Here, we perform a generalized scaling analysis of the magnetization data.  We infer the presence of a phase-mediated transition from the fitting relation $\sqrt{M} \propto (T_a(H)-T)^m$, where $T_a(H)$ denotes the apparent transition temperature, and $m$ is the fitting exponent.  For underdoped samples in particular, the phase fluctuations may be quite large, and the exponent $m$ may vary significantly from its mean-field value. 

To focus on phase-mediated behavior, we must distinguish it from the complementary fluctuations of the amplitude of the order parameter, which appear as anomalous enhancements of the magnetization above $T_c$, particularly for the field orientation $H\parallel \hat{c}$.  Like the phase fluctuations \cite{beck}, these amplitude fluctuations occur predominantly in the pseudo-gap phase of underdoped materials, due to their extreme anisotropy or two-dimensionality \cite{norman}.  Indeed, deoxygenated crystals with same transition temperatures studied here have previously been shown to exhibit strong 2D critical fluctuations \cite{alvarenga}.  The phase-mediated behavior of interest here occurs at temperatures below the amplitude fluctuation regime, in the vicinity of the crossing point.  For our analysis, we determine a temperature fitting range for each data set.  For the curves obtained with $H\parallel \hat{c}$, we take the inflection point of the magnetization as our high temperature cutoff, since it marks a change in the temperature dependence (curvature) of the magnetization.  Typically, the inflection point remains within the superconducting scaling regime, since it lies within a degree or two below the zero-field transition.  For most of the curves the inflection point occur at the crossing point. For the curves with $H\parallel ab$ the crossing point is not well defined (neither a change in curvature), and the high temperature cutoff was taken visually from the smoothness of the asymptotic behavior of each curve near $T_c$. For the low temperature cutoff, we choose a point where the reversible magnetization becomes approximately linear.  The resulting fitting curves are shown as solid lines in Figs.~1-3. Dashed lines in Figs.~1-3 (only for $H\parallel \hat{c}$) represents  the extrapolation of the fittings to $T_a(H)$, and also helps to visualize the change in the curvature in each curve. We note that the fitting ranges have a typical width of 5-8~K, which tends to increase for higher fields with $H\parallel \hat{c}$, and to decrease for more underdoped samples.  For the narrower fitting ranges, we included some magnetization data lying in the linear $M(T)$ regime, corresponding to Abrikosov's solution of the Ginzburg-Landau equations \cite{abrikosov}.  Such linear data have previously been used to estimate mean-field transition temperatures $T_c(H)$ from isofield $M(T)$ data \cite{welp}.  Figure ~4(a) illustrates the linear region of the reversible magnetization curves for selected fields and the respective $T_c(H)$ obtained from the linear extrapolation. Figures ~4(b) and ~4(c) show plots of  $\sqrt{M}$  vs. $T$ (and fittings) obtained from two $M$ vs. $T$ curves  of Fig. ~4(a), evidencing the differences between $T_c(H)$ and  $T_a(H)$ for these samples. We mention that values of $T_c(H)$ obtained from this linear method produced consistent values of $dH_{c2}/dT<0$ for all samples in reasonable agreement with values listed in Ref.~\onlinecite{JLTP1}. Results for the fitted values of $T_a(H)$ and $m$ are shown for different samples and field orientations in Figs.~1-3.  We note that most of the resulting fitting curves can be extended for several degrees below their fitting range (up to 5 K for high field curves), while still providing a good description of the experimental data.  For the Nb sample in the inset of Fig. \ref{92K}(a), the resulting values of $T_a(H)$ are essentially identical to the mean-field transition $T_c(H)$ obtained in Ref.~\onlinecite{saidnb}, with $(dT_a(H)/dH)<0$.  For the optimally doped YBaCuO sample in  Fig. \ref{92K}(b), we also obtain transition temperatures consistent with the linear $M(T)$ extrapolation method used in Ref.~\onlinecite{welp}, for both field orientations.  One may expect this conventional behavior for both of these samples, which exhibit no pseudo-gap phase.  We note that the observed fitting exponents for optimally doped YBaCuO were enhanced slightly for $H\parallel \hat{c}$, and suppressed slightly for $H\parallel ab$, relative to their mean-field values.  However, these deviations are relatively weak, compared with the underdoped samples.

The underdoped samples in Figs.~2 and 3 show behavior that is not characteristically mean-field, and reflect enhanced fluctuation effects.  For $H\perp \hat{c}$, $T_a(H)$ is essentially constant, with no resolvable slope.  However, for $H\parallel c$, we observe a distinct, anomalous slope with $dT_a(H)/dH>0$.  In fact, for high field curves, the inflection points occur above the respectives crossing points, but below $T_c$, as shown in the plots of Fig.~4(a) and 4(b), which partially explains why $T_a(H)$ increases with field.  For these curves, the fitting exponents $m$ also appear to increase anomalously with the field. 

\section{Discussion}
\begin{figure}[t]
\includegraphics[width=\linewidth]{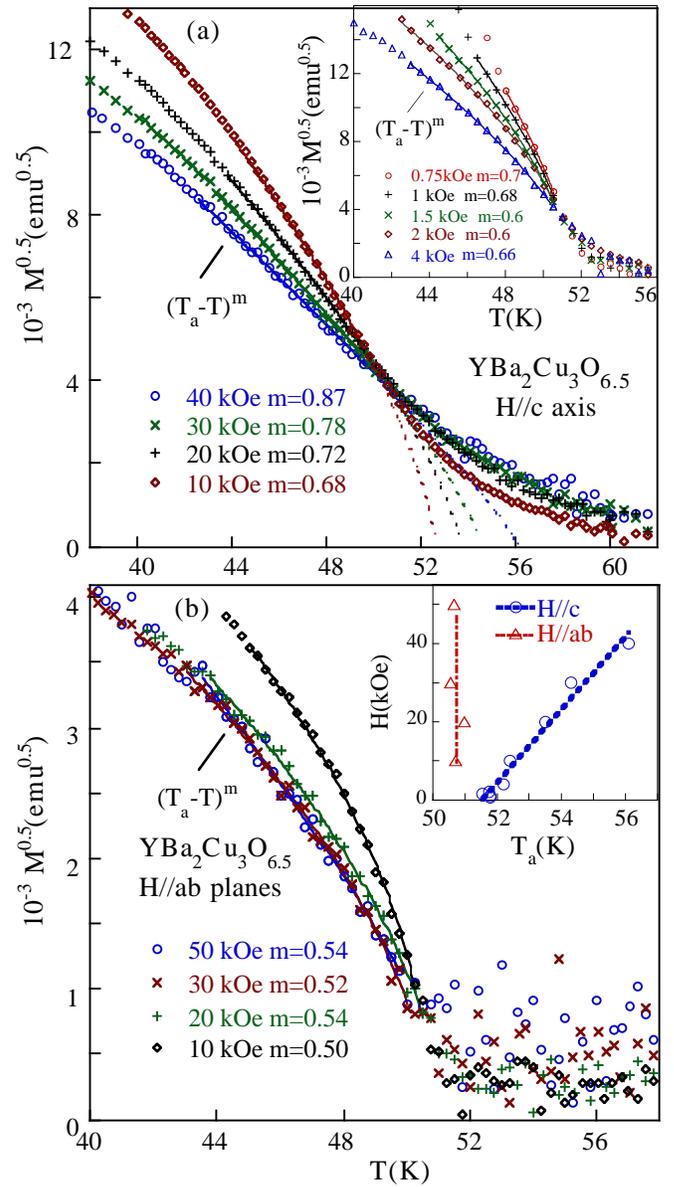}
\caption{Isofield curves of $\sqrt{M}$ vs. $T$ for underdoped YBaCuO ($T_c$=52~K), where $M$ is the reversible magnetization.  (a) $H\parallel \hat{c}$, higher fields. Dashed lines shown the extrapolation of the fittings to $T_a(H)$.  Inset:  lower fields. (b) Same sample, $H\parallel ab$.  Inset:  phase correlation temperature $T_a(H)$ vs.\ $H$.}
\label{52K}
\end{figure}

\begin{figure}[t]
\includegraphics[width=\linewidth]{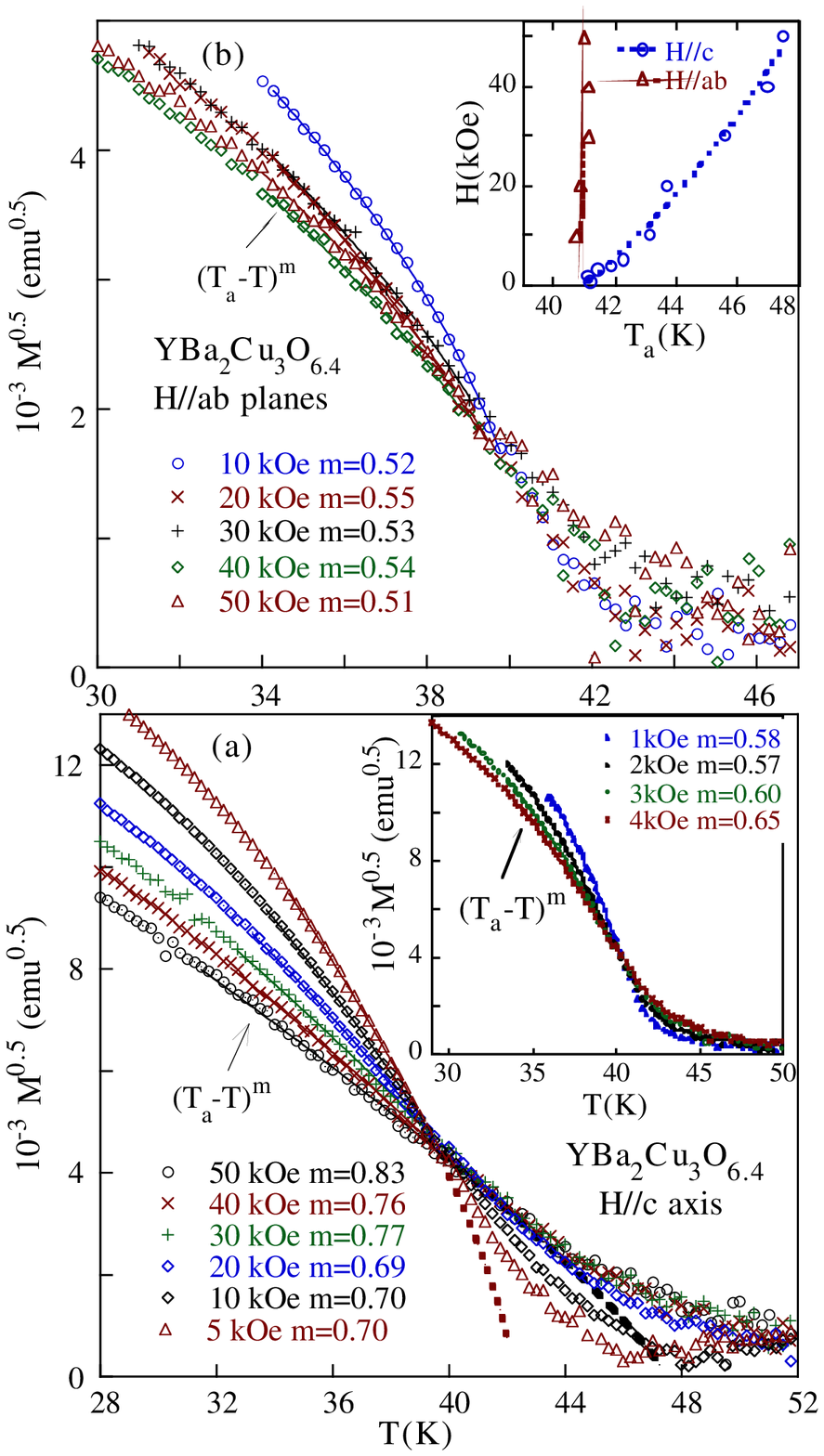}
\caption{Isofield curves of $\sqrt{M}$ vs. $T$ for underdoped YBaCuO ($T_c$=41.5~K), where $M$ is the reversible magnetization.  (a) $H\parallel \hat{c}$, higher fields. Dashed  lines shown the extrapolation of the fittings to $T_a(H)$ for $H=$ 50 and 5 kOe.  Inset:  lower fields. (b) Same sample, $H\parallel ab$.  Inset:  phase correlation temperature $T_a(H)$ vs.\ $H$.}
\label{41K}
\end{figure}

It is shown in Ref.~\onlinecite{kwon} that phase fluctuations of the order parameter with d-wave symmetry can have a net effect on the superfluid density of states reducing the gap in the vicinity of $T_c$. An inspection of Figure 2 of Ref.~\onlinecite{kwon} shows that the change in the shape of the gap near $T_c$ is accompanied by a change in the value of the exponent $m$. We observed that we can apply the later result to explain the large values of the fitting exponent $m$ found for $YBaCuO$ with $H\parallel c$.  Following this conjecture the temperature dependence of the order parameter in the curves of Fig. \ref{92K}, \ref{52K}, and \ref{41K} is defined by phase fluctuations, and values of $T_a(H)$ obtained from the fittings represents temperature values at which phase coherence is lost.  Then the temperature dependence of the magnetization observed here in several different samples is generally consistent with the proposed theories of Curty and Beck \cite{beck} and Emery and Kivelson \cite{emery}.  In this picture, the inferred transition temperature $T_a(H)$ represents the onset of  phase coherence, and may be associated with the zero-field temperature $T_\phi$.  Although $T_a(H)$ does not correspond to a true phase transition, it appears to control the scaling properties of the magnetization $M(T)$ in the asymptotic regime, just below the regime of strong amplitude fluctuations.  For conventional superconductors like Nb, phase and amplitude fluctuations are both weak, and the magnetization exhibits mean-field behavior.  For optimally doped YBaCuO phase fluctuations are expected to be important in the vicinity of $T_c$ \cite{emery}, but the absence of a pseudo-gap phase produced  a mean-field like behavior for $T_a(H)$ as for Nb.  However, underdoped high-$T_c$ superconductors exhibit very different behavior, with significant phase fluctuations, and a line of phase-mediated transitions $T_a(H)$ quite distinct from the mean-field transition $T_c(H)$.   The observation of phase coherence above $T_c$ correlates with the presence of a pseudo-gap in the underdoped materials \cite{beck}.   The persistence of phase coherence above $T_c(H)$ (up to $T_a(H)$) may explain why fluctuation effects observed above $T_c$ in underdoped materials resemble superconducting properties normally observed below $T_c$, as claimed in Refs.~\onlinecite{wang1,wang2,said}].  The distinction between $T_a(H)$ and $T_c(H)$ may also explain why the observed fitting exponent deviates from $m=1/2$.  Since doping and magnetic field both effect phase fluctuations \cite{klemm}, we can anticipate that the value of $m$ will depend on the field.

A possible explanation for the anomalous behavior of $dT_a/dH$ and $m$ in the underdoped samples arises from the $d$-wave pairing symmetry of the order parameter.  As mentioned above, it is shown in  Ref.~\onlinecite{kwon} that phase fluctuations of a $d$-wave order parameter can have a net effect on the superfluid density of states  producing a change in the value of the fitting exponent $m$.  In this picture, the anisotropic dependence of $m$ on $H$ can also be explained in terms of the order parameter symmetry.  Differences on the effect of phase fluctuations depending on the symmetry, s-wave or d-wave, of the order parameter are expected since the d-wave symmetry presents a node and anti-node where the effects of phase and amplitude fluctuations are distinct \cite{kwon}. This fact suggests that  differences observed in the values of the exponent $m$ depending on the direction of the magnetic field with respect to the c-axis are consequences of the pairing symmetry of the order parameter in YBaCuO, which is likely to be d-wave in the ab-planes and s-wave along the c-axis (as in Nb).  We note that this pairing symetry of the order parameter have been suggested by J. Mannhart et al. \cite{mannhart} for high-$T_c$ superconductors and a recent result in a LaSrCuO crystal also suggests $s$-wave pairing along the $\hat{c}$-axis \cite{muller}.  

\begin{figure}[t]
\includegraphics[width=\linewidth]{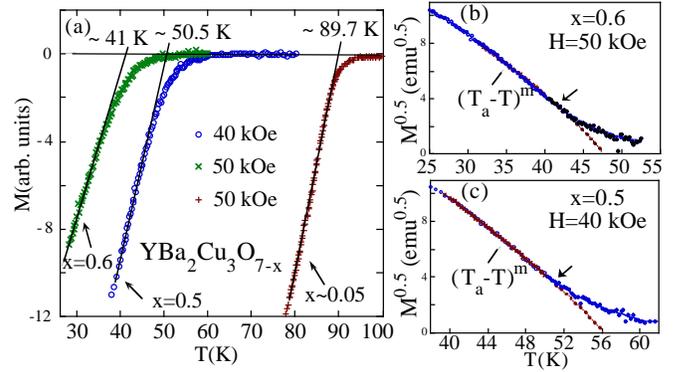}
\caption{(a) Linear extrapolation of the reversible magnetization for YBaCuO isofield curves with $H\parallel \hat{c}$. (b) and (c)  $\sqrt{M}$ vs. $T$ for two $M$ vs. $T$ curves of Fig.~4a ($T_c$= 41.5 and 52 K) . Dashed lines  represents the  extrapolation of the fittings to $T_a(H)$. The arrow in each curve indicates the position of the inflection point.}
\label{fig4}
\end{figure}

An alternative explanation for the anistropic fitting exponent $m$ may arise from the anisotropy of the vortex critical current in the quasi-2D, underdoped cuprates \cite{referee}.  For these highly anisotropic superconductors, the high-field vortex state is formed of pancake vortices when $H\parallel \hat{c}$ and Josephson vortices when $H\perp \hat{c}$, with dissipation that is field independent for the case of $H\parallel ab$ \cite{glazman}.  The zero-field-cooled  (reverslble) magnetization may therefore provide a direct measurement of the anisotropy of the critical current, which sustain the vortex state,  and its temperature dependence.  We point out, however, that the observed behavior is strongly influenced by fluctuations, leading to complications for interpretations based on mean-field arguments.

In conclusion, we have determined the apparent transition temperature $T_a(H)$, associated with phase coherence above the mean-field transition, by fitting the temperature dependence of the magnetization $\sqrt{M} \propto [T_a(H)-T]^m$ for several high-$T_c$ and conventional superconductors.  For deoxygenated YBaCuO samples with $H\parallel \hat{c}$ we observe that $T_a(H)>T_c$ with $dT_a/dH>0$.  Such anomalous behaviour suggests that the magnetization is probing a regime where  phase coherence persist above $T_c$, even in the absence of true long-range order.  These same samples also exhibit significant amplitude fluctuations without  phase coherence extending to well above $T_a(H)$.  For all our samples, the values of the fitting exponent $m\simeq 0.5$ are consistent with mean-field theory when $H\parallel ab$, but deviate from this conventional behavior when $H\parallel c$ axis, particularly for the underdoped cuprates.  We have suggested several interpretations for this anomalous behavior, including an anisotropy of the symmetry of the order parameter:  $s$-wave along the $\hat{c}$-axis vs.\ $d$-wave in the $ab$ planes.

\begin{acknowledgements}
We thank Mark Friesen for a critical reading of the manuscript and  many helpful suggestions, Boyd Veal who kindly provide the YBaCuO  crystals and J.C. Campuzano for helpful discussions. ADA acknowledges support  from CNPq. 
\end{acknowledgements}


\begin{thebibliography}{99}
\bibitem{referee}This interpretation was pointed out by the referee.
\end{thebibliography}
\end{document}